\documentclass[twocolumn,showpacs,pra,superscriptaddress]{revtex4}
\usepackage{dcolumn}
\usepackage{graphicx}
\usepackage{bbm}

\newcommand{\be}{\begin{equation}}
\newcommand{\ee}{\end{equation}}
\newcommand{\beq}{\begin{eqnarray}}
\newcommand{\eeq}{\end{eqnarray}}

\newcommand{\hide}[1]{}

\newcommand{\ket}[1]{\left| #1 \right\rangle}

\begin{document}

\title{Enhanced index of refraction in four-wave mixing media}

\author{Elena Kuznetsova}
\affiliation{Department of Physics, University of Connecticut,
Storrs, CT 06269}
\affiliation{ITAMP, Harvard-Smithsonian Center
for Astrophysics, Cambridge, MA 02138} 
\affiliation{Russian Quantum Center, 100 Novaya Street, Skolkovo, Moscow region, 143025, Russia}
\author{Renuka Rajapakse}
\affiliation{Department of Physics, University of Connecticut,
Storrs, CT 06269}
\author{S. F. Yelin}
\affiliation{Department of Physics, University of Connecticut,
Storrs, CT 06269} 
\affiliation{ITAMP, Harvard-Smithsonian Center
for Astrophysics, Cambridge, MA 02138}
\affiliation{Department of Physics, Harvard University, Cambridge, MA 02138}
\date{\today}

\begin{abstract}
Refractive index enhancement accompanied by vanishing absorption in a four-level atomic system interacting with two control and 
two probe fields in a regime of four-wave mixing (FWM) has been predicted and studied 
in the present work. We analyzed the maximal index enhancement in the four-level FWM system and gave index estimates for a real atomic gas of $^{40}$K, 
taking into account its multilevel structure and collisional and Doppler broadenings at large atomic densities. We also discussed how vanishing 
absorption with no nearby amplification can be realized in a two species system, consisting of a four-level FWM and a two-level system, where the latter provides 
additional absorption for one of the 
probe fields. We numerically estimated the index change in a system composed of $^{40}$K and $^{39}$K gases. 
\end{abstract}

\maketitle 

\section{Introduction}

Refractive index governs light propagation in macroscopic materials, resulting in everyday phenomena such as light refraction, 
reflection and absorption \cite{Principles-of-optics}. 
For a long time the index of refraction was considered given by nature and could be changed only by choosing an appropriate material. 
In recent decades this situation has changed in two different ways: 1) control of material optical response using external fields 
and 2) artificial materials (metamaterials) have been realized. One of the best examples of the first kind is electromagnetically induced 
transparency (EIT), which reduces absorption of a resonant probe field 
by irradiating a medium with a strong control field thus making an otherwise opaque medium 
transparent \cite{EIT-review}. It also changes dramatically the group velocity of light, reducing it down to cm/s or even stopping a light pulse 
completely by converting it 
into a material excitation \cite{Slow-light,Stopped-light}. Control of the group velocity 
is important for many applications such as optical delay lines and memories, but the refractive index itself is also of great interest. 
For example, negative index of refraction, proposed by Veselago \cite{Veselago}, can provide imaging with 
unlimited resolution allowing one to build a perfect lens \cite{Pendry-lens}. On the opposite side, enhanced refractive index attracts significant 
attention as well. 
It is well known that spatial resolution of optical imaging techniques is given by the focus size of a light wave, which is diffraction limited to 
$\lambda^{2}$. Overcoming this limit and reducing the minimal image feature size is one of the main goals in spectroscopy nowadays \cite{Principles-of-nanooptics}. 
Enhanced refractive index leads to a decreased wavelength of an electromagnetic wave 
in a medium $\lambda=\omega/cn$, where $n$ is the real part of the complex refractive index. 
Thus light with shorter wavelengths in high index media can find applications in lithography and optical imaging. Another exciting 
application of enhanced refractive index is invisibility cloacking, allowing to hide 
an object from light by tailoring the material index around it \cite{Invis-cloacking}.

Refractive index can be controlled using external fields in analogy to group velocity. It is well known that in a medium composed of two-level 
atoms the index of refraction 
for a near resonant light can be high, but absorption, given by the imaginary part of the 
complex index, is of the same order. It results in light attenuation by $1/e$ at a distance corresponding to the 
accumulated phase $\sim 1$ rad, making this approach impractical. 
In order to eliminate absorption and have at the same time enhanced index of refraction two schemes have been 
suggested for laser controlled atomic media. The first one pioneered by Scully \cite{Scully-enh-ind} 
utilizes a long-lived coherence at a low-frequency transition in a three-level $\Lambda$-type atomic system. The coherence modifies the 
medium susceptibility and allows one  
to cancel probe absorption at some frequency, having at the same time non-vanishing refractive index. The second approach, proposed by 
Yavuz \cite{Yavuz-enh-ind}, does not require atomic coherence and uses a superposition of two-level absorbing and amplifying resonances, 
shown in Fig.\ref{fig:Suscept-two-level}a.
There is a frequency corresponding to vanishing absorption between the resonances accompanied by non-zero refractive index. 
The main difficulty of both approaches is that there is an amplification region next to the zero absorption point and if the field frequency fluctuates 
around this point the 
fluctuations get amplified.
To overcome this problem it was proposed in \cite{Kochar} to use two overlapping resonances of different width: a wide absorbing and a narrow amplifying one. 
In this case vanishing absorption with no nearby amplification can be realized (see Fig.\ref{fig:Suscept-two-level}d). 

The remaining difficulty of the absorbing/amplifying system of \cite{Yavuz-enh-ind,Kochar} is the need for either direct or Raman inversion 
at the amplifying transition. Direct inversion in two-level systems is difficult to realize and mantain. Raman inversion 
 in three-level species is easier to maintain but requires first pumping population in one of the two ground states. In the scheme used in \cite{Yavuz-exper1} 
where absorbing and amplifying resonances were realized with $^{87}$Rb and $^{85}$Rb gases, cross-coupling between optical pumping 
processes in the two species limited index enhancement to $\Delta n \approx 2\times 10^{-7}$. 
In a more optimal case of \cite{Yavuz-exper2} where the absorbing and amplifying transitions 
were realized in a single atomic species, population had to be optically pumped into a single ground state sublevel. In this case optical 
pumping worked well for atomic densities $N<1.2\cdot 10^{14}$ cm$^{-3}$, also limiting index enhancement to $\Delta n\sim 10^{-4}$. 
The index change obtained using the atomic coherence based approach of Scully \cite{Scully-experim} was also limited to $\Delta n\sim 10^{-4}$ due to 
 population pumping into dark hyperfine states reducing the number of atoms actively interacting with light.

We propose a more robust scheme for realization of enhanced refractive index based on a four-wave mixing (FWM) system, shown in 
Fig.\ref{fig:FWM-scheme}a. In our approach an atomic coherence provides the enhanced index combined with vanishing absorption similar to \cite{Scully-enh-ind}.   
The FWM scheme, however, can be realized without any kind of optical pumping, inversion or population transfer in dark states making it easier and more robust compared to both the coherence based 
and the absorbing/amplifying systems.

The paper is organized as follows. In Section IIa we remind the index enhancement approach based on two-level atomic species providing absorbing 
and amplifying resonances for a probe field. In Section IIb we analyze the complex refractive index for a medium composed of four-level atoms in the 
regime of four-wave mixing. In Section IIc we consider the possibility to realize index enhancement with vanishing absorption and no amplification 
regions nearby. Finally, we conclude in Section III.

\section{Index enhancement in an absorbing/amplifying medium}

\begin{figure}[h]
\center{
\includegraphics[width=3.in]{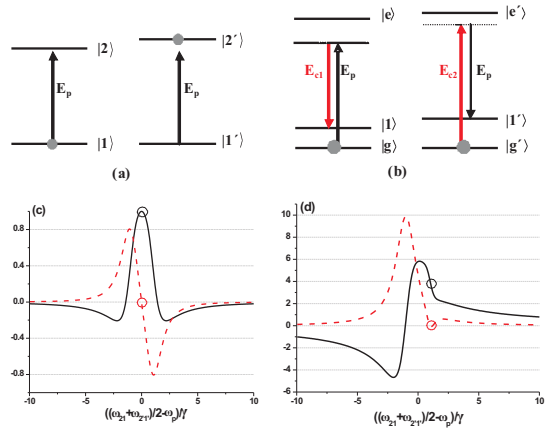}
\caption{\label{fig:Suscept-two-level} (Color online) a) Two-component medium of two-level absorbing and amplifying systems; b) Two-component 
 medium of three-level systems interacting with control and probe fields in a Raman configuration. The probe field experiences Raman absorption and 
amplification by the first and second system, respectively; 
c) Real (dashed line) $\Delta n'/(2\pi N |\mu_{21}|^{2}/\hbar)$ and imaginary (solid line) $\Delta n''/(2\pi N |\mu_{21}|^{2}/\hbar)$ parts of the complex refractive index for equal widths $\gamma=\gamma'$ of the absorbing and amplifying resonances, 
$N=N'$, $\Delta-\Delta'=2\gamma$ and $\rho_{11}-\rho_{22}=-(\rho_{11}'-\rho_{22}')=1$. Frequencies of 
vanishing absorption accompanied by enhanced refractive index are shown by circles; d) Real (dashed line) and imaginary (solid line)
parts of the complex refractive index for a wider absorbing resonance with $\gamma'=0.5\gamma$, $N'=0.1N$, $\Delta-\Delta'=2\gamma$ and $\rho_{11}-\rho_{22}=-(\rho_{11}'-\rho_{22}')=1$.}
}
\end{figure}

In the approach of \cite{Yavuz-enh-ind,Kochar} a two-component atomic medium is used, 
in which one species provides absorption for a 
probe field while the other species provides amplification. Absorption and amplification can be realized either with two-level 
(see Fig.\ref{fig:Suscept-two-level}a) or three-level Raman systems. 
In practice it is difficult to realize and maintain inversion at optical transitions, and therefore Raman transitions are used. 
The equivalent of Fig.\ref{fig:Suscept-two-level}a for Raman transitions is shown in Fig.\ref{fig:Suscept-two-level}b, 
which uses two Raman systems with all population in the ground states $\ket{g}$, $\ket{g'}$, driven by two strong control fields $E_{c1}$, $E_{c2}$ and 
interacting with the same weak probe field $E_{p}$. The probe and the control fields are near two-photon resonant with low 
frequency transitions $\ket{g}-\ket{1}$, $\ket{g'}-\ket{1'}$ and the probe field is absorbed at the $\ket{g}-\ket{e}$ and 
amplified at the $\ket{e'}-\ket{1'}$ transitions. This scheme can be realized using either two atomic species as in Fig.\ref{fig:Suscept-two-level}b 
or different transitions in the same atomic species as was done in the experiment \cite{Yavuz-exper2}. 
The change of the refractive index induced by the absorbing/amplifying system of Fig.\ref{fig:Suscept-two-level}a, which we consider for simplicity, is given by
\begin{eqnarray}
\Delta n=\frac{2\pi i}{\hbar}\left(\frac{N|\mu_{21}|^{2}(\rho_{11}-\rho_{22})}{i\Delta+\gamma}+ \right. \nonumber \\
\left. +\frac{N'|\mu_{2'1'}|^{2}(\rho_{11}'-\rho_{22}')}{i\Delta'+\gamma'}\right),
\end{eqnarray}
where $N$, $N'$ are atomic densities of the first and second species, $\mu_{21}$, $\mu_{2'1'}$ are the dipole moments of the 
corresponding transitions, $\Delta$, $\Delta'$ are the detunings of the probe field from the $\ket{2}-\ket{1}$ and $\ket{2'}-\ket{1'}$ 
transitions, and $\gamma$ and $\gamma'$ are the optical coherence decay rates. 

When $\rho_{11}-\rho_{22}>0$ and $\rho_{1'1'}-\rho_{2'2'}<0$ the imaginary part of the refractive index $\Delta n''$ will be 
a superposition of absorbing and amplifying resonances, and $\Delta n'$ will show two frequency regions with normal and anomal 
dispersion shifted with respect to each other. Depending on the frequency shift between the resonances it allows one to have enhanced or reduced refractive index with vanishing absorption 
shown by circles in Fig.\ref{fig:Suscept-two-level}c. Vanishing absorption is accompanied in this case by amplification on the 
right side, and if the frequency fluctuates it will result in amplification of frequency fluctuations shifting the frequency to smaller 
index values. If the absorbing resonance is 
wider than the amplifying one and the energy difference between the two resonances is less that the 
 width of the wider resonance \cite{Kochar}, it becomes possible to realize enhanced (or reduced) refractive index combined with zero absorption 
and no nearby gain. 
The corresponding real and imaginary parts of the refractive index are shown in Fig.\ref{fig:Suscept-two-level}d. 

We mentioned in the introduction that the difficulty of the two-level absorbing/amplifying 
scheme is the need for inversion for the amplifying two-level system. The frequencies of the absorbing and amplifying two-level 
systems are close and the pump, providing inversion, can degrade the population in the absorbing system. This problem can be avoided if 
three-level Raman systems are used, but in this case  
population has to be transferred into a specific ground state by optical pumping. As was observed in \cite{Yavuz-exper2} 
optical pumping works efficiently only up to certain atomic density, which was $N\sim 1.2 \cdot 10^{14}$ cm$^{-3}$ in the case of $^{87}$Rb. 
For higher densities, resulting in higher refractive index, optical pumping cannot populate a single ground state sublevel and degrades the system performance. 
In the next section we describe how the absorbing/amplifying system can be replaced by a single species four-level four-wave mixing 
system, which allows to have enhanced index with vanishing absorption without the need for inversion and optical pumping.

\section{Susceptibility of a four-level FWM system}

\begin{figure}[h]
\center{
\includegraphics[width=3.2in]{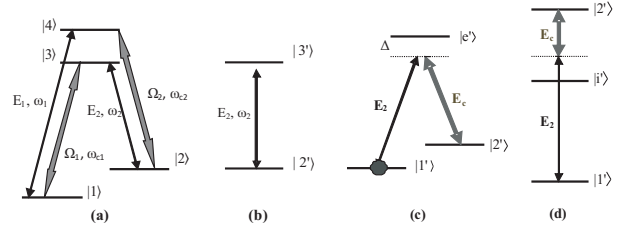}
\caption{\label{fig:FWM-scheme} a) Four-level four-wave mixing atomic system with two control fields (Rabi frequencies $\Omega_{1}$ and $\Omega_{2}$) 
close to resonance with the $\ket{3}-\ket{1}$ and $\ket{4}-\ket{2}$ transitions. The other transitions $\ket{4}-\ket{1}$ and $\ket{3}-\ket{2}$ interact 
with two probe fields $E_{1}$ and $E_{2}$ having frequencies $\omega_{1}$ and $\omega_{2}$; b) Absorbing two-level system providing 
vanishing probe absorption with no amplification on both sides; 
c), d) Additional absorption for one of the probe fields, e.g. for $E_{2}$, can be realized via Raman or two-photon absorption using an 
additional control field $E_{c}$.}
}
\end{figure}

We consider a four-level four-wave mixing scheme shown in Fig.\ref{fig:FWM-scheme}a. In the FWM medium there are two strong control fields at frequencies 
$\omega_{1c}$ and $\omega_{2c}$, which we assume to have constant 
Rabi frequencies $\Omega_{1}$ and $\Omega_{2}$, and two weak probe fields with different frequencies $\omega_{1}$ and $\omega_{2}$:
\begin{equation}
E_{j}=Re\left({\cal E}_{j}e^{-i\omega_{j}t+ik_{j}z}\right),\; j=1,2,
\end{equation}
where ${\cal E}_{1}$ and ${\cal E}_{2}$ are the field amplitudes. The density matrix equation for the system interacting with the fields is  
$\frac{d\hat{\rho}}{dt}=\frac{i}{\hbar}\left[\hat{\rho},\hat{H}\right]+{\cal L}\hat{\rho}$, where $\hat{H}=\hat{H}_{0}+\hat{H}_{int}$ and 
\begin{eqnarray}
\hat{H}_{0}=\hbar\omega_{41}|4\rangle \langle 4|+\hbar\omega_{31}|3\rangle \langle 3|+\hbar\omega_{21}|2\rangle \langle 2|, \nonumber \\
\hat{H}_{int} = -\vec{\mu}_{41}\vec{E}|4\rangle \langle 1|-\vec{\mu}_{32}\vec{E}|3 \rangle \langle 2|-\vec{\mu}_{31}\vec{E} |3 \rangle \langle 1| \nonumber \\
-\vec{\mu}_{42}\vec{E}|4 \rangle \langle 2|+H.c.,
\end{eqnarray}
$\vec{E}=\vec{E}_{c1}+\vec{E}_{c2}+\vec{E}_{1}+\vec{E}_{2}$ and ${\cal L}\hat{\rho}$ is the Lindblad operator describing population and coherence decay.  
Elements of the density matrix obey the following equations:
\begin{eqnarray}
\label{eq:density-matrix}
\frac{d\sigma_{41}}{dt} & = & -(i(\omega_{41}-\omega_{1})+\gamma_{41})\sigma_{41}-i\alpha_{1}(\rho_{44}-\rho_{11}) \nonumber \\
&& -i\Omega_{1}\sigma_{43}+i\Omega_{2}\sigma_{21}, \nonumber \\
\frac{d\sigma_{32}}{dt} & = & -(i(\omega_{32}-\omega_{2})+\gamma_{32})\sigma_{32}-i\alpha_{2}(\rho_{33}-\rho_{22})+ \nonumber \\
&& +i\Omega_{1}\sigma_{21}^{*}e^{-i\Delta kz}-i\Omega_{2}\sigma_{43}^{*}e^{-i\Delta kz}, \nonumber \\
\frac{d\sigma_{21}}{dt} & = & -(i(\omega_{21}-\omega_{1}+\omega_{2})+\gamma_{21})\sigma_{21}-i\Omega_{1}\sigma_{32}^{*}e^{-i\Delta kz} \nonumber \\
&& -i\alpha_{1}\sigma_{42}^{*}+i\alpha_{2}^{*}\sigma_{31}e^{-i\Delta kz}+i\Omega_{2}^{*}\sigma_{41}, \nonumber \\
\frac{d\sigma_{43}}{dt} & = & -(i(\omega_{43}-\omega_{1}+\omega_{c1})+\gamma_{43})\sigma_{43}-i\Omega_{1}^{*}\sigma_{41} \nonumber \\
&& -i\alpha_{2}^{*}\sigma_{42}e^{-i\Delta kz}+i\alpha_{1}\sigma_{31}^{*}+i\Omega_{2}\sigma_{32}^{*}e^{-i\Delta kz}, \nonumber \\
\frac{d\sigma_{31}}{dt} & = & -(i(\omega_{31}-\omega_{c1})+\gamma_{31})\sigma_{31}-i\Omega_{1}(\rho_{33}-\rho_{11}) \nonumber \\
&& -i\alpha_{1}\sigma_{43}^{*}+i\alpha_{2}\sigma_{21}e^{i\Delta kz}, \nonumber \\
\frac{d\sigma_{42}}{dt} & = & -(i(\omega_{42}-\omega_{c2})+\gamma_{42})\sigma_{42}-i\Omega_{2}(\rho_{44}-\rho_{22}) \nonumber \\
&& -i\alpha_{2}\sigma_{43}e^{i\Delta kz}+i\alpha_{1}\sigma_{21}^{*}, \nonumber \\
\frac{d\rho_{11}}{dt} & = & -i\Omega_{1}\sigma_{31}^{*}+i\Omega_{1}^{*}\sigma_{31}+\Gamma_{31}\rho_{33}+\Gamma_{41}\rho_{44}, \nonumber \\
\frac{d\rho_{22}}{dt} & = & -i\Omega_{2}\sigma_{42}^{*}+i\Omega_{2}^{*}\sigma_{42}+\Gamma_{32}\rho_{33}+\Gamma_{42}\rho_{44}, \nonumber \\
\frac{d\rho_{33}}{dt} & = & i\Omega_{1}\sigma_{31}^{*}-i\Omega_{1}^{*}\sigma_{31}-(\Gamma_{31}+\Gamma_{32})\rho_{33}, \nonumber \\
\rho_{44} & = & 1-\rho_{11}-\rho_{22}-\rho_{33}, 
\end{eqnarray}
where $\sigma_{41}=\rho_{41}e^{i\omega_{1}t-ik_{1}z}$, $\sigma_{32}=\rho_{32}e^{i\omega_{2}t-ik_{2}z}$, $\sigma_{31}=\rho_{31}e^{i\omega_{c1}t-ik_{c1}z}$, 
$\sigma_{42}=\rho_{42}e^{i\omega_{c2}t-ik_{c2}z}$, $\sigma_{21}=\rho_{21}e^{i(\omega_{1}-\omega_{c2})t-i(k_{1}-k_{c2})z}$, 
$\sigma_{43}=\rho_{43}e^{i(\omega_{1}-\omega_{c1})t-i(k_{1}-k_{c1})z}$, $\alpha_{1}=\mu_{41}{\cal E}_{1}/2\hbar$, $\alpha_{2}=\mu_{32}{\cal E}_{2}/2\hbar$ are 
Rabi frequencies of the probe and $\Omega_{1}=\mu_{31}E_{c1}/2\hbar$, $\Omega_{2}=\mu_{42}E_{c2}/2\hbar$ are Rabi frequencies of the control fields, 
and $\Delta k=(\vec{k}_{1}-\vec{k}_{c1}+\vec{k}_{2}-\vec{k}_{c2})_{z}$ is the wavevector mismatch of the four fields. We assume the fields to propagate in $z$ 
direction. Coherence and population decay rates are denoted as 
$\gamma_{ij}$ and $\Gamma_{ij}$, respectively.

The medium polarization $P_{j}=Re\left({\cal P}_{j}e^{-i\omega_{j}t+ik_{j}z}\right),\; j=1,2$ at the frequencies $\omega_{1}$ and $\omega_{2}$ of 
the probe fields has an amplitude 
\begin{eqnarray}
{\cal P}_{1}=N\mu_{14}\sigma_{41}, \nonumber \\
{\cal P}_{2}=N\mu_{23}\sigma_{32},
\end{eqnarray}
and is given by \cite{Misha-FWM}:
\begin{eqnarray}
{\cal P}_{1}=\chi_{11}{\cal E}_{1}+\chi_{12}e^{i\delta k z}{\cal E}_{2}^{*}, \nonumber \\
{\cal P}_{2}=\chi_{22}{\cal E}_{2}+\chi_{21}e^{i\delta k z}{\cal E}_{1}^{*},
\end{eqnarray} 
where $\chi_{ij}$ are the medium susceptibilities.

To simplify the analysis we assume that: 1) the control fields have equal Rabi frequencies $\Omega_{1}=\Omega_{2}=\Omega$ and 
are resonant with their corresponding transitions: $\omega_{42}-\omega_{2}=0$, $\omega_{31}-\omega_{3}=0$. 
Due to frequency matching condition $\omega_{1}-\omega_{c1}=\omega_{c2}-\omega_{2}$ the probe fields then have detunings 
$\omega_{41}-\omega_{1}=\delta$, $\omega_{32}-\omega_{2}=-\delta$; 3) decay rates of excited states 
$\ket{3}$ and $\ket{4}$ are equal $\Gamma_{3}=\Gamma_{4}=\Gamma_{r}$ and $\Gamma_{ij}=\Gamma_{r}/2$, where $i=3,4$ and $j=1,2$, 
and $\Gamma_{r}$ is the excited state radiative decay rate. Optical coherence decay rates are also assumed equal 
$\gamma_{ij}=\gamma$ for i=3,4 and j=1,2. 
Decay of the low-frequency coherence at the $\ket{2}-\ket{1}$ transition is assumed to be 
much slower than at optical transitions $\gamma_{21} \ll \gamma$. 
The corresponding susceptibilities are then given by \cite{Misha-FWM} 
\begin{eqnarray}
\label{eq:suscepts}
\chi_{11} & = & \frac{i|\mu_{41}|^{2}N}{2\hbar}(\rho_{ll}-\rho_{uu})\frac{i\delta +\gamma_{21}-i\delta\Omega^{2}/(\gamma(\gamma+i\delta))}{(i\delta+\gamma)(i\delta+\gamma_{21})+2|\Omega|^{2}}, \nonumber \\
\chi_{12} & = & -i\frac{\mu_{41}\mu_{32}^{*}N}{2\hbar\gamma(i\delta+\gamma)}\frac{\Omega^{2}(\rho_{ll}-\rho_{uu})(i\delta+2\gamma)}{(i\delta+\gamma)(i\delta+\gamma_{21})+2|\Omega|^{2}},  \\
\chi_{22} & = & \frac{i|\mu_{32}|^{2}N}{2\hbar}(\rho_{ll}-\rho_{uu})\frac{-i\delta +\gamma_{21}+i\delta\Omega^{2}/(\gamma(\gamma-i\delta))}{(-i\delta+\gamma)(-i\delta+\gamma_{21})+2|\Omega|^{2}},  \nonumber \\
\chi_{21} & = & -i\frac{\mu_{41}^{*}\mu_{32}N}{2\hbar\gamma(-i\delta+\gamma)}\frac{\Omega^{2}(\rho_{ll}-\rho_{uu})(-i\delta+2\gamma)}{(-i\delta+\gamma)(-i\delta+\gamma_{21})+2|\Omega|^{2}},  \nonumber
\end{eqnarray}
where the population difference between 
the ground and excited states, assuming for simplicity equal dipole moments of optical transitions $\mu_{41}=\mu_{42}=\mu_{32}=\mu_{31}=\mu$ is given by
\begin{equation}
\rho_{ll}-\rho_{uu}=\frac{1}{2}\frac{1}{1+4\Omega^{2}/\gamma \Gamma_{r}}.
\end{equation}

As a result, propagation equations for the two probe fields look as follows:
\begin{eqnarray}
\label{eq:prop-eqs}
\frac{\partial {\cal E}_{1}}{\partial z}=2\pi i k_{1} \chi_{11}{\cal E}_{1}+2\pi i k_{1}\chi_{12}e^{i\Delta k z}{\cal E}_{2}^{*}, \nonumber \\
\frac{\partial {\cal E}_{2}}{\partial z}=2\pi i k_{2} \chi_{22}{\cal E}_{2}+2\pi i k_{2}\chi_{21}e^{i\Delta k z}{\cal E}_{1}^{*}. 
\end{eqnarray}

\begin{figure}[h]
\center{
\includegraphics[width=3.5in]{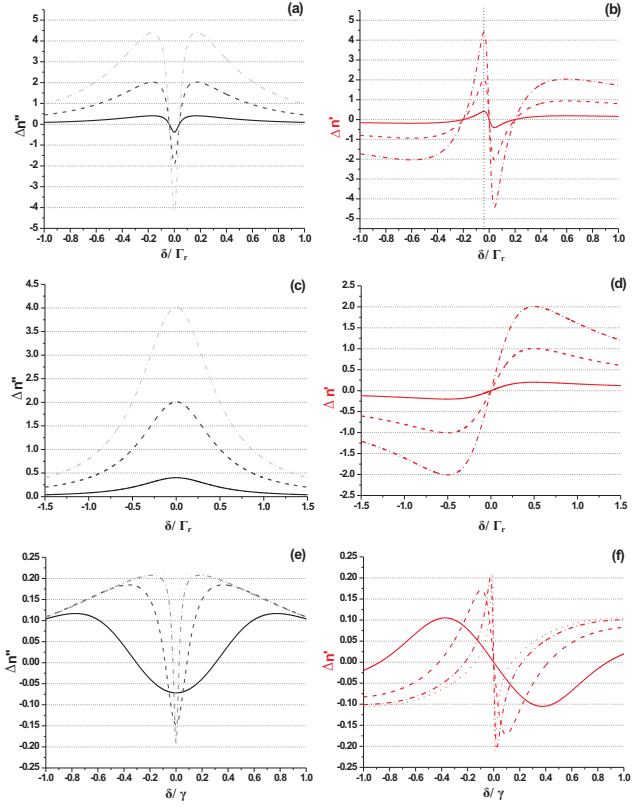}
\caption{\label{fig:suscept-no-abs} (Color online) a) Imaginary (grey lines) and b) real (red lines) parts of the refractive index change $\Delta n$, 
corresponding to the $\lambda_{+}$ propagation 
constant. Curves are shown for a range of atomic densities given by the parameter 
$r=\pi |\mu|^{2}N/2\hbar \Gamma_{r}=3N\lambda^{3}/64\pi^{2}$. Solid, dashed and dash-dotted lines correspond to 
$N=10^{14},\;5\cdot 10^{14}$ and $10^{15}$ cm$^{-3}$. Other parameters are: $\lambda=770.1$ nm, $\gamma=\Gamma_{r}/2$, 
$\gamma_{21}=10^{-3}\Gamma_{r}$ and $\Omega=0.1\Gamma_{r}$.  
The vertical dotted line denotes the frequency of maximal refractive index with vanishing absorption; 
c) Imaginary and d) real parts of the refractive index 
corresponding to $\lambda_{-}$ for the same densities; e),f) Dependence of index enhancement on the low-frequency coherence decay rate $\gamma_{21}$:  
e) imaginary (grey lines) and f) real (red lines) 
parts of the index change corresponding to $\lambda_{+}$ for a range of decay rates $\gamma_{21}$. 
Solid, dashed, dash-dotted and dotted lines correspond to 
$\gamma_{21}=0.1\gamma$, $0.01\gamma$, $0.001\gamma$ and $0.0001\gamma$; $N=10^{14}$ cm$^{-3}$ and $\Omega$ maximizing $\Delta n'$ 
according to Eq.(\ref{eq:F-equation}) was used.}
}
\end{figure}

Setting for simplicity $\Delta k=0$ we can find from Eqs.(\ref{eq:prop-eqs}) propagation constants for the probe fields:
\begin{eqnarray}
\label{eq:prop-constants}
\lambda^{\pm}=i\pi(k_{1}\chi_{11}-k_{22}\chi_{22}^{*}) \pm \nonumber \\
\pm i\pi \sqrt{(k_{1}\chi_{11}+k_{22}\chi_{22}^{*})^{2}-4k_{1}k_{2}\chi_{12}\chi_{21}^{*}}.
\end{eqnarray}
Assuming $k_{1}\approx k_{2}=k$ (e.g. states $\ket{1}$ and $\ket{2}$ and similarly $\ket{3}$ and $\ket{4}$ are hyperfine sublevels of the same 
electronic level) and using $\chi_{22}=-\chi_{11}^{*}$ and $\chi_{21}=-\chi_{12}^{*}$ we have 
\begin{equation}
\label{eq:index}
\lambda^{\pm}=2\pi i k(\chi_{11}\pm \chi_{12}).
\end{equation}
The change in the refractive index $\Delta n=\Delta n'+i\Delta n''$ induced by the FWM medium can be found from the propagation constants as 
$\Delta n'_{\pm}={\rm Im} \lambda_{\pm}/k$ and $\Delta n''_{\pm}=-{\rm Re} \lambda_{\pm}/k$.
Typical shapes of the imaginary and real parts of the refractive index $\Delta n''$ and $\Delta n'$ in this idealized four-level medium 
are shown in Figs.\ref{fig:suscept-no-abs}a and \ref{fig:suscept-no-abs}b. The index is calculated for a range of atomic densities $N$ described by a dimensionless parameter 
$r=\pi |\mu|^{2}N/\hbar \Gamma_{r}$. Taking into account that $\Gamma_{r}=4k^{3}|\mu|^{2}/3\hbar$ gives $r=3N\lambda^{3}/64\pi^{2}$, where 
$\lambda=2\pi/k$, which determines the maximal possible index enhancement. One can see that for $\lambda_{+}$ there 
are regions of enhanced and reduced index accompanied by zero absorption (and amplification). The frequency corresponding to vanishing absorption, 
which at the same time corresponds to maximal refractive index, is denoted by a vertical dotted line. The $\lambda_{-}$ constant is always 
accompanied by non-zero absorption, as shown in Fig.\ref{fig:suscept-no-abs}c. The enhanced refractive index in the FWM system is realized without 
inversion or optical pumping, populations are assumed to be redistributed by the control fields.

\begin{figure}[h]
\center{
\includegraphics[width=2.2in]{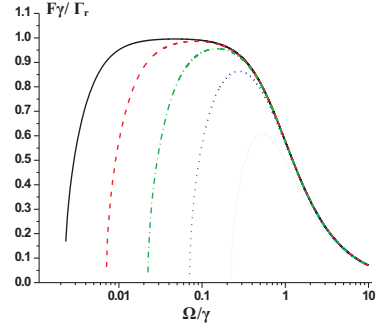}
\caption{\label{fig:maxn-FWM} (Color online) Dependence of the $F$ coefficient given by Eq.(\ref{eq:F-equation}) on the Rabi frequency of the control fields for a range of 
low-frequency coherence decay rates $\gamma_{21}$. Increasing $F$ corresponds to $\gamma_{21}=10^{-1}\gamma$ (light blue short-dotted curve), $10^{-2}\gamma$ (blue dotted curve), $10^{-3}\gamma$ (green dash-dotted curve), 
$10^{-4}\gamma$ (red dashed curve) and $10^{-5}\gamma$ (black solid curve), where $\gamma=\Gamma_{r}/2$. }
}
\end{figure}

We can estimate the maximal possible index enhancement in the idealized FWM system. From Eq.(\ref{eq:index}) the complex index $\Delta n_{+}=\lambda_{+}/ik$ equals
\begin{equation}
\Delta n_{+}=\frac{i\pi N|\mu|^{2}}{\hbar}(\rho_{ll}-\rho_{uu})\frac{i\delta +\gamma_{21}-2\Omega^{2}/\gamma}{(i\delta+\gamma)(i\delta+\gamma_{21})+2\Omega^{2}}.
\end{equation}
Absorption vanishes when the imaginary part of the index $\Delta n_{+}''=0$ at $\delta^{2}=\frac{4\Omega^{4}/\gamma^{2}-\gamma_{21}^{2}}{1+2\Omega^{2}/\gamma^{2}}$. 
The corresponding real part of the index at this frequency is given by $\Delta n_{+}'=rF$, where
\begin{eqnarray}
\label{eq:F-equation}
F=\frac{\Gamma_{r}/\gamma}{\left(1+4\Omega^{2}/\gamma \Gamma_{r}\right)}\sqrt{\frac{4\Omega^{4}/\gamma^{4}-\gamma_{21}^{2}/\gamma^{2}}{1+2\Omega^{2}/\gamma^{2}}}\times \nonumber \\
\times \frac{\left(2\left(1+\Omega^{2}/\gamma^{2}\right)-\gamma_{21}/\gamma\right)\left(1+2\Omega^2/\gamma^2\right)}{\left(1+\gamma_{21}/\gamma\right)\left(4\Omega^{2}(1+\Omega^{2}/\gamma^{2})/\gamma^{2}+\gamma_{21}\left(2-\gamma_{21}/\gamma\right)/\gamma\right)}.
\end{eqnarray}
The factor $F$ is maximized for some Rabi frequency of the control fields, the corresponding dependence is shown in 
Fig.\ref{fig:maxn-FWM} for a range of low-frequency coherence decay rates $\gamma_{21}$. It shows that 
for small enough $\gamma_{21}\le 10^{-3}\gamma$ the maximal value of $F\sim \Gamma_{r}/\gamma$. It allows us to estimate the maximal possible 
enhancement of the 
refractive index in the four-level FWM medium: $\Delta n_{+\;max}'=r\Gamma_{r}/\gamma\approx 3 N\Gamma_{r}\lambda^{3}/64 \gamma \pi^{2}$. Using as an example 
a D1 transition of $^{40}$K with $\lambda=770.1$ nm we get $\Delta n_{+\;max}'=2.17\cdot 10^{-15}\Gamma_{r} N(cm^{-3})/\gamma$. 
We obtain therefore that for radiatively broadened optical transitions with $\gamma=\Gamma_{r}/2$ (where $\Gamma_{r}=6.035$ MHz for the D1 line) 
 $\Delta n_{+}'\approx 0.43$ already for $N=10^{14}$ cm$^{-3}$ and even larger index change can be realized for higher densities.

\section{Index enhancement in the presence of collisional and Doppler broadening}

\begin{figure*}[h]
\center{
\includegraphics[width=6.in]{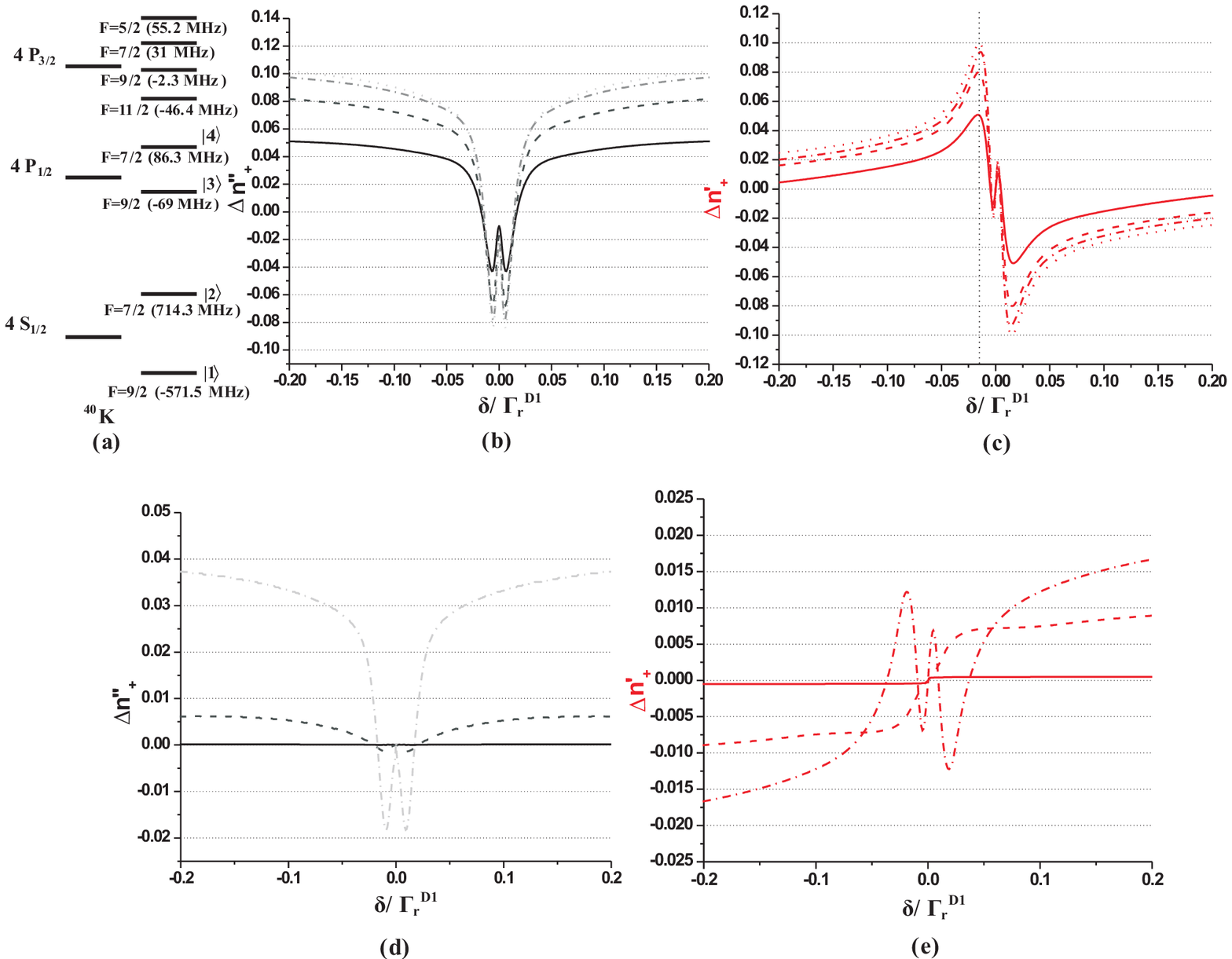}
\caption{\label{fig:Coll-broad} (Color online) a) Hyperfine structure of the 4 S$_{1/2}$, 4 P$_{1/2}$ and 4 P$_{3/2}$ states of $^{40}$K, including level shifts 
relative to centers of D1 and D2 transitions. Hyperfine states forming the FWM scheme are also shown; 
b) Imaginary (grey lines) and c) real (red lines) parts of the complex index change for a four-level scheme formed by hyperfine 
transitions of $^{40}$K D1 line, taking into account collisional broadening. Solid, dashed, dash-dotted and dotted lines correspond to the 
density of K vapor is $N=10^{14}$, $5\cdot 10^{14}$, 
$10^{15}$ cm$^{-3}$ and $5\cdot 10^{15}$ cm$^{-3}$. The frequency corresponding to vanishing absorption and enhanced index is shown by a vertical dotted line; 
d) Imaginary (grey lines) and e) real (red lines) parts of the complex index change for a four-level scheme formed by hyperfine 
transitions of the $^{40}$K D1 and D2 lines (see text) taking into account collisional and Doppler broadenings. Solid, dashed, and dash-dotted 
lines correspond to the temperature of K vapor $T=300$, $450$ and $600$ K.}
}
\end{figure*}

Large enhancement of refractive index ($\Delta n'\sim 4$ in Fig.\ref{fig:suscept-no-abs}a) obtained in the previous section assumes 
radiatively broadened optical transitions. 
For large atomic densities, however, collisional broadening has to be taken into account adding a contribution $\gamma_{coll}$ to the optical 
coherence decay rate $\gamma=\Gamma_{r}/2+\gamma_{coll}$. In order to get a more realistic estimate of index enhancement we will analyze the D1 transition of $^{40}$K 
taking into account its multilevel structure. The ground 4 S$_{1/2}$ and excited 4 P$_{1/2}$ states of $^{40}$K are each split 
in two hyperfine sublevels with $F=9/2,\;7/2$ ($F=9/2$ state being lower in energy)
due to the nuclear spin $I=4$ of $^{40}$K. The hyperfine splittings are $1.286$ GHz in the ground and $0.155$ GHz in the excited state 
\cite{K-properties}, as shown in Fig.\ref{fig:Coll-broad}a. 
Four levels of the FWM scheme can be associated with the hyperfine states as
$\ket{1}=\ket{4\;S_{1/2},F=9/2}$, $\ket{2}=\ket{4\;S_{1/2},F=7/2}$ and 
$\ket{3}=\ket{4\;P_{1/2},F'=9/2}$, $\ket{4}=\ket{4\;P_{1/2},F'=7/2}$. We assume that the control and probe fields are $\pi$ polarized, 
coupling all $\ket{F,m_{F}}$ ground state to all $\ket{F',m_{F}}$ excited state hyperfine sublevels, respectively.

The fractional hyperfine 
quantum numbers of $^{40}$K
make all transitions $\ket{F,m_{F}}\leftrightarrow \ket{F',m_{F}}$ allowed, which does not happen for integer hyperfine numbers in the case of 
e.g. Rb and Cs. In the latter 
case the $\ket{F,m_{F}=0}\leftrightarrow \ket{F'=F,m_{F}'=0}$ transition is forbidden, resulting in optical pumping of all population into the 
dark $\ket{F,m_{F}=0}$ state and vanishing light-matter interaction. One can try to use circularly polarized fields to overcome this difficulty, 
but in this case all population will be optically pumped into another dark state $\ket{F,m_{F}=F}$ and light-matter interaction vanishes again. 
Therefore, to counteract optical pumping into dark states in systems with integer hyperfine numbers some sort of repumping has to be used. 
Fractional hyperfine quantum numbers of $^{40}$K allow one to avoid pumping population into dark states, and as a result no repumping is needed.

We can now estimate the index enhancement in the four-level system of the previous section at large atomic densities taking into account collisional broadening of optical transitions. 
The broadening rate for the D1 transition of K is $2\gamma_{coll}=2\beta N=0.7\cdot 10^{-13}N(cm^{-3})$ MHz 
according to \cite{K-coll-broadening}. Maximal possible refractive index enhancement is then given by  
$\Delta n_{+\;max}'=2.17\cdot 10^{-15}\Gamma_{r} N/(\Gamma_{r}/2+\beta N)$, resulting in theoretical maximal value 
$\Delta n_{+\;max}' \approx 2.17\cdot 10^{-15}\Gamma_{r}/\beta \approx 0.37$ for 
$N\gg \Gamma_{r}/2\beta=8.6 \cdot 10^{13}$ cm$^{-3}$.

In order to get a better estimate for index enhancement we considered interaction of the $\pi$ polarized control and probe fields with the real 
multilevel structure of the $^{40}$K D1 line, taking into account transition dipole moments and population decay rates from \cite{K-properties}.  The 
index change in the presence of collisional broadening is shown in Figs.\ref{fig:Coll-broad}b and \ref{fig:Coll-broad}c. One can see that in the real system the maximal index change is 
reduced with respect to the theoretical estimate to $\Delta n'_{+}\approx 0.1$, but is still high.

So far we have considered the situation which corresponds to cold gases with vanishing Doppler 
broadening. In warm and hot atomic vapors Doppler broadening has to be taken into account. We included its effect by numerically averaging 
propagation constants (\ref{eq:prop-constants}) over the 1D Maxwell-Boltzmann distribution 
$f(\delta \omega)=\exp(-(\delta \omega/W_{D})^{2})/\sqrt{\pi}W_{D}$, where $\delta \omega=kv_{z}$ is the Doppler frequency shift of an atom having 
velocity $v_{z}$ along the propagation axis z; $W_{D}=k\sqrt{2k_{B}T/m}$ is the width of the Doppler profile. Due to a small mass of $^{40}$K the Doppler 
width is large even at room temerature ($W_{D}=458$ MHz at $T=300$ K), exceeding the hyperfine splitting in the 4 P$_{1/2}$ state. 
It will result in interaction of all fields with both $\ket{3}=\ket{F'=9/2}$ and $\ket{4}=\ket{F'=7/2}$ states, complicating the analysis. 
To avoid this complication we assumed that the state $\ket{4}=\ket{F'=9/2}$ belongs to the 4 P$_{3/2}$ excited state, i.e. to the D2 line. We again 
took into account 
dipole moments and population decay rates for transitions between $\ket{F,m_{F}}$ and $\ket{F'=9/2,m_{F}}$ of the D2 line from \cite{K-properties}.

The temperature 
not only controls the Doppler width but vapor density as well, making it temperature dependent. The pressure-temperature and density-pressure 
dependence for K vapor 
is the following \cite{K-properties}
\begin{eqnarray}
\log_{10}p=7.9667-\frac{4646}{T}, \;\; 298\;K\;<T<336.8\;K,\nonumber \\
\log_{10}p=7.4077-\frac{4453}{T}, \;\; 336.8\;K<T<600\;K, \nonumber \\
N=\frac{10^{2}p}{k_{B}T},
\end{eqnarray}
where $p$ is in mbar, $T$ in K and $N$ is in m$^{-3}$.
The resulting complex index $\Delta n_{+}$ corresponding to the 
$\lambda_{+}$ propagation constant is given in Figs.\ref{fig:Coll-broad}b and \ref{fig:Coll-broad}c for several temperatures. One can see that for high temperatures 
$T\sim 600$ K the index change is $\Delta n_{+}'\approx 0.01$, i.e. about an order of magnitude smaller than in a cold gas without Doppler broadening. 

Finally, we discuss briefly how the probe fields propagate in a medium of finite thickness. From 
Fig.\ref{fig:suscept-no-abs} we know that the imaginary part of $\Delta n_{-}$ is positive, 
i.e. corresponds to absorption for all detunings. As a result, the $\lambda_{-}$ component will be attenuated during propagation, while the 
$\lambda_{+}$ component will get amplified or stay constant at the frequency of vanishing absorption. Eqs.(\ref{eq:prop-eqs}) can be solved to obtain the following finite thickness solutions for the fields:
\begin{eqnarray}
\label{eq:prop-fields}
E_{1}(z=L)=e^{i\pi \delta a L}\left[E_{1}(z=0)\left(i(k_{1}\chi_{11}+k_{2}\chi_{22}^{*})\sin(\pi SL)+ \right. \right. \nonumber \\
\left. \left. +S\cos(\pi SL)\right)/S+E_{2}^{*}(z=0)2ik_{1}\chi_{12}\sin(\pi SL)/S \right], \nonumber \\
E_{2}^{*}(z=L)=e^{i\pi \delta a L}\left[-2ik_{2}\chi_{21}^{*}\sin(\pi SL)E_{1}(z=0)/S+ \right.  \\
\left. +E_{2}^{*}(z=0)\left(\cos(\pi SL)-i(k_{1}\chi_{11}+k_{2}\chi_{22}^{*})\sin(\pi SL)/S\right) \right], \nonumber 
\end{eqnarray}
where $\delta a=k_{1}\chi_{11}-k_{2}\chi_{22}^{*}$, $S=\sqrt{\left(k_{1}\chi_{11}+k_{2}\chi_{22}^{*}\right)^{2}-4k_{1}k_{2}\chi_{12}\chi_{21}^{*}}$, 
$L$ is the medium thickness, $E_{1,2}(z=0)$ are the field amplitudes at the medium entrance. At a distance $|Im \lambda_{-}|L \gg 1$ Eqs.(\ref{eq:prop-fields}) 
have the form:
\begin{eqnarray} 
E_{1}(z=L)\sim \frac{1}{2S}e^{i\lambda_{+}L}\left(E_{1}(z=0)(k_{1}\chi_{11}+k_{2}\chi_{22}^{*}+S)+ \right. \nonumber \\
\left. +2k_{1}\chi_{12}E_{2}^{*}(z=0)\right), \nonumber \\
E_{2}^{*}(z=L)\sim \frac{1}{2S}e^{i\lambda_{+}L}\left(-2k_{2}\chi_{21}^{*}E_{1}(z=0)+ \right. \nonumber \\
\left. +E_{2}^{*}(z=0)(S-k_{1}\chi_{11}-k_{2}\chi_{22}^{*})\right), \nonumber 
\end{eqnarray}
which shows that it suffices to have a single field at the medium entrance, e.g. the $E_{1}(z=0)$ field, to realize index enhancement. The second field 
will be generated during propagation due to four-wave mixing process.

We also note that the four-wave mixing system can have another interesting application. It can be used to realize parity-time (PT) symmetric complex index of refraction obeying the condition
$n(x,t)=n^{*}(-x,-t)$, i.e. with real and imaginary parts of the index being even and odd functions in time and space, respectively. 
The PT symmetric refractive index allows to model non-Hermitian Hamiltonians of the Schrodinger equation having real eigenvalues, related to 
discussions on fundamentals of quantum mechanics based on axioms of Hermitian operators for observables.
The complex 
index derived from propagation constants (\ref{eq:prop-constants}) has the required symmetry with respect to frequency inversion $\omega \rightarrow -\omega$, 
which is equivalent to time inversion $t \rightarrow -t$. The symmetries of $n'$ and $n''$ with respect to inversion in space can be realized using far-detuned 
control fields propagating in $x$ direction producing spatially dependent Stark shifts \cite{PT-with-atoms}.

\section{Composite FWM and two-level absorbing system}

In the previous section we showed that enhanced refractive index accompanied by vanishing absorption can be realized in a four-level atomic system 
in the regime of four-wave mixing. However, one can see from Figs.\ref{fig:suscept-no-abs} and \ref{fig:Coll-broad} that there is amplification on the right side of 
the vanishing absorption point, which will amplify frequency components of the probe pulse in this region.   
Vanishing absorption with no nearby amplification is possible if an additional absorbing atomic component is added to the system. Absorption for one of the probe fields, e.g. for 
the $E_{2}$ field, can be realized by a two-level system shown in 
Fig.\ref{fig:FWM-scheme}b. In this case the total susceptibility at the $\omega_{2}$ frequency is  
$\chi_{22}=\chi_{22}^{FWM}+\chi_{22}^{abs}$, where $\chi_{22}^{FWM}$ is given by Eq.(\ref{eq:suscepts}) and the contribution to susceptibility from the 
two-level system is 
\begin{eqnarray}
\label{eq:two-level-suscept}
\chi_{22}^{abs}=i\frac{|\mu_{32}'|^{2}N'}{\hbar}\frac{1}{i(\omega_{32}'-\omega_{2})+\gamma_{32}'},
\end{eqnarray}
where the detuning $\omega_{32}'-\omega_{2}=-\delta+\delta_{0}$, $\delta_{0}=\omega_{32}'-\omega_{32}$ is the frequency difference of the 
transitions in the two and four-level systems, and $\delta=-(\omega_{32}-\omega_{2})$ is the detuning for the four-level system.

\begin{figure*}[h]
\center{
\includegraphics[width=6.in]{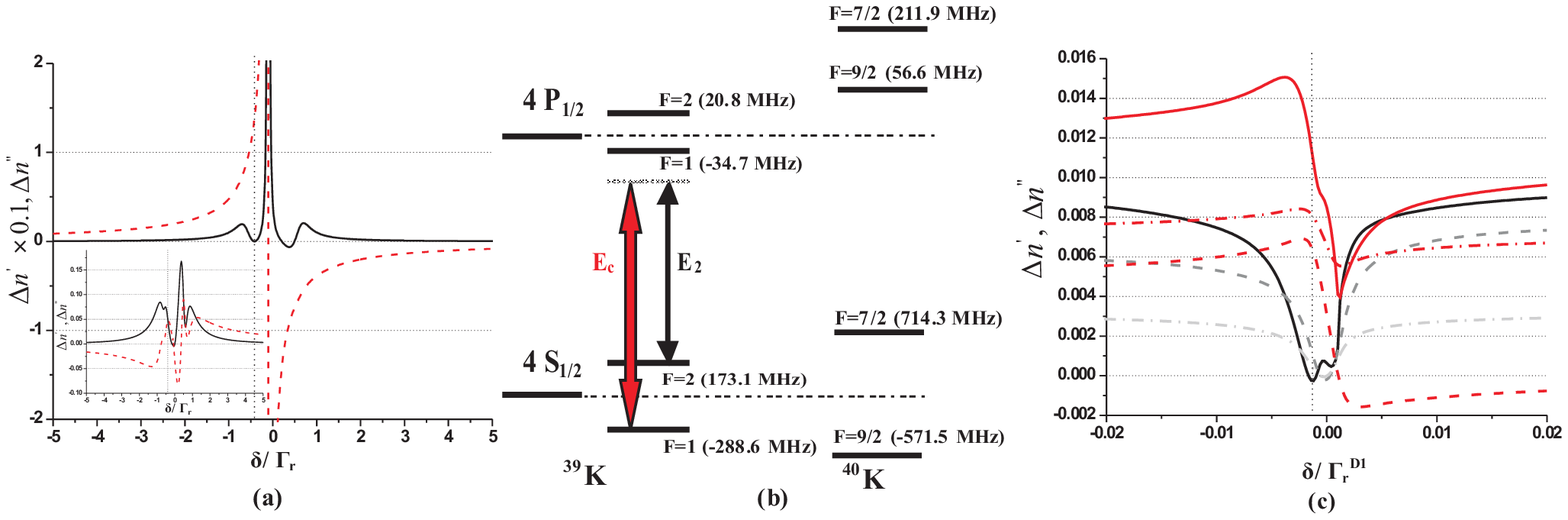}
\caption{\label{fig:composite-system} (Color online) a) Real (red dashed line) and imaginary (black solid line) parts of the complex index change for the $\lambda_{+}$ 
constant of the composite four and two-level system. Ideal four-level FWM plus a two-level absorbing systems are assumed with the 
susceptibilities given by Eqs.(\ref{eq:suscepts}), (\ref{eq:two-level-suscept}). The frequency corresponding to vanishing absorption with no 
nearby amplification is shown by a vertical dotted line. 
Parameters are: $N_{FWM}=10^{14}$ cm$^{-3}$, $N_{abs}=10N_{FWM}$, $\gamma=\Gamma_{r}/2$, $\gamma_{abs}=10^{-3}\Gamma_{r}/2$, 
$\delta_{0}=0.1\Gamma_{r}$, $\Omega=0.45\Gamma_{r}$, where $\Gamma_{r}$ is the radiative decay rate of excited states in the four-level system; 
b) Hyperfine structure of D1 transitions of $^{39}$K and $^{40}$K including level shifts relative to transition center. Shown also 
the Raman transition for $E_{2}$ and an additional control field 
near resonant with $F=1\leftrightarrow F=2$ of $^{39}$K providing absorption for the probe field;  
c) Real (red lines) and imaginary (grey lines) parts of the complex index change for $\lambda_{+}$ propagation constant in the  
$^{39}$K+$^{40}$K system. Solid lines correspond to $N_{40K}=10^{14}$ cm$^{-3}$, $N_{39K}=10N_{40K}$, $\Omega_{c1}=\Omega_{c2}=1.3\Gamma_{r}^{D1}$, 
$\Omega_{c}=10\Gamma_{r}^{D1}$, $\delta_{0}=-0.92\Gamma_{r}^{D1}$, $\gamma_{9/2,7/2}=\gamma_{21}=10^{-3}\Gamma_{r}^{D1}$; dashed lines correspond 
to $N_{40K}=5\cdot 10^{14}$ cm$^{-3}$, $N_{39K}=15N_{40K}$, $\Omega_{c1}=\Omega_{c2}=2.3\Gamma_{r}^{D1}$, 
$\Omega_{c}=20\Gamma_{r}^{D1}$, $\delta_{0}=-3\Gamma_{r}^{D1}$, $\gamma_{9/2,7/2}=\gamma_{21}=10^{-3}\Gamma_{r}^{D1}$; 
short-dashed lines correspond to $N_{40K}=10^{15}$ cm$^{-3}$, $N_{39K}=25N_{40K}$, $\Omega_{c1}=\Omega_{c2}=4.4\Gamma_{r}^{D1}$, 
$\Omega_{c}=20\Gamma_{r}^{D1}$, $\delta_{0}=-3\Gamma_{r}^{D1}$, $\gamma_{9/2,7/2}=\gamma_{21}=10^{-3}\Gamma_{r}^{D1}$, where $\Gamma_{r}^{D1}$ 
is the radiative decay rate of the D1 transition of $^{40}$K.}
}
\end{figure*}

The propagation constants for the composite system can be calculated from Eq.(\ref{eq:prop-constants}) using the modified $\chi_{22}$.
Fig.\ref{fig:composite-system}a shows the corresponding real and imaginary parts of the complex index corresponding to $\lambda_{\pm}$ 
assuming the idealized radiatively broadened four-level system described by Eqs.(\ref{eq:suscepts}) and a 
radiatively broadened two-level system. One can see that for the $\lambda_{+}$ constant there is vanishing 
absorption at $\delta/\Gamma_{r} \approx -0.42$ accompanied by enhanced refractive index, shown by the vertical line, with no amplification on either sides. 
The index change corresponding to $\lambda_{-}$, shown in the inset to Fig.\ref{fig:composite-system}a, displays strong absorption at the 
frequency of vanishing absorption of $\lambda_{+}$.

The absorbing system can also be realized by a Raman or two-photon transition shown in Fig.\ref{fig:FWM-scheme}c. The advantage of the Raman and 
two-photon configurations is that the optical transition frequency does not have to be near resonant with the probe field. In this case 
resonant absorption for the probe field will be provided by the two-photon resonance of the probe and a control field $E_{c}$ with the transition $\ket{2'}-\ket{1'}$. 
An additional and important advantage of the Raman scheme is that the resonance can be quite narrow, much narrower than the 
optical one, which we found necessary to have vanishing absorption with no nearby amplification in Fig.\ref{fig:composite-system}.

To have a more realistic estimate of the index enhancement in the composite case we considered a two-species medium of $^{40}$K and $^{39}$K atomic 
gases. The $^{40}$K isotope 
provides the four-level system analyzed in the previous section, and $^{39}$K allows to realize absorption for one of the probe fields, specifically, for the 
$E_{2}$ field. The two isotopes of which $^{39}$K is more abundant (93.26$\%$) than $^{40}$K (0.01$\%$) have the same D1 transition wavelength 
$\lambda=770.1$ nm, but different hyperfine structure, due to the $^{39}$K nuclear spin of $3/2$. 
The corresponding level scheme is shown in Fig.\ref{fig:composite-system}b. One can see that resonant absorption for $E_{2}$ can be 
realized in a Raman scheme where the probe and an additional control field are near two-photon resonant with the $F=1\leftrightarrow F=2$ hyperfine 
transition of 4 S$_{1/2}$ state of $^{39}$K. In numerical calculations we assumed that the control field has the same $\pi$ polarization as the 
probe one and took into account fields interaction with all $\ket{F,m_{F}}\leftrightarrow \ket{F',m_{F}}$ transitions, 
where $F=1,2$ and $F'=1,2$ are the hyperfine sublevels of the 
4 S$_{1/2}$ and 4 P$_{1/2}$ states of $^{39}$K. We also took into account collisional broadening of the optical transitions of both isotopes 
using the total atomic density $N=N_{40K}+N_{39K}$. 
The complex index change corresponding to the $\lambda_{+}$ propagation constant is 
shown in Fig.\ref{fig:composite-system}c.  The refractive index at the vanishing absorption point is enhanced by $\Delta n_{+}'\sim 10^{-2}$ in the case 
of $N_{40K}=10^{14}$ cm$^{-3}$ and $N_{39K}=10N_{40K}$, shown by the vertical line.

\section{Conclusions}

We analyzed a possibility to realize enhanced refractive index with vanishing absorption in a four-level four-wave mixing atomic system.  
The susceptibility of the FWM system displays frequency regions where absorption turns into amplification accompanied by non-zero refractive index, 
leading to enhanced refractive index with zero absorption in this medium. 
Index enhancement in the FWM system is more robust compared to previously considered approaches using a three-level coherence-based $\Lambda$ scheme 
and a composite absorbing/amplifying scheme  
because it allows one to avoid pumping population into dark states and does not require inversion or pumping population into specific states. 
Although optical pumping of population into dark states does not take place in ideal three or two-level 
schemes, it becomes an issue in real multilevel atomic systems, e.g. in alkali gases.
We analyzed 
a particular system of $^{40}$K gas, which does not have this problem, and found that refractive index enhancement $\Delta n\sim 0.1$ is possible in a cold gas with 
vanishing Doppler broadening at densities $N\sim 10^{15}$ cm$^{-3}$. The index change is $\Delta n\sim 10^{-2}$ if both collisional and Doppler broadening are 
present. We note that these predictions give larger index change compared to observed experimentally so far by two orders of magnitude. 

We showed also that enhanced refractive index accompanied by vanishing absorption with no 
nearby amplification can be realized in a composite system including four-level four-wave mixing and 
two-level absorbing species. As an example we considered a combination of $^{40}$K and $^{39}$K atomic gases, where the first system provides 
FWM and the second one provides additional absorption for one of the probe fields. Analysis of the complex index of the composite system taking 
into account interaction of the fields with multilevel atomic structures and collisional broadening showed that index enhancement $\Delta n\sim 10^{-2}$ can be 
realized at a frequency of vanishing absorption with no nearby amplification.

The authors gratefully acknowledge financial support from the Russian Quantum Center and NSF.

\end{document}